\documentclass[final]{aipproc}%
\usepackage[english]{babel}
\usepackage{amsmath}
\usepackage{amsfonts}
\usepackage{amssymb}
\usepackage{graphicx}%
\providecommand{\U}[1]{\protect\rule{.1in}{.1in}}
\providecommand{\U}[1]{\protect\rule{.1in}{.1in}}
\layoutstyle{8x11double}
\begin{document}

\title{Cosmological singularity}


\author{V.Belinski}
{address={ICRANet, p.le della Repubblica, 10 - 65122 Pescara, Italy} }

\begin{abstract}
The talk at international conference in honor of Ya. B.
Zeldovich 95th Anniversary, Minsk, Belarus, April 2009. The talk represents
a review of the old results and contemporary development on the problem of cosmological singularity.
\end{abstract}

\maketitle

\section{Preface}

This problem was born in 1922 when A.Friedmann wrote his famous cosmological
solution for the homogeneous isotropic Universe. However, during the next 35
years researches devoted their attention mainly to the physical processes
after the Big Bang and there was no serious attempts to put under a rigorous
analysis the phenomenon of the cosmological singularity as such. The person
who inspired the beginning of such analysis was L.D.Landau. In the late 1950s
he formulated the crucial question whether the cosmological singularity is a
general phenomenon of General Relativity or it appears only in particular
solutions under the special symmetry conditions. The large amount of work have
been done in Landau school before an answer emerges in 1969 in the form of the
so-called ``BKL conjecture'' (the present-day terminology). The basic
reviews\footnote{Some of these articles should not be considered merely as
reviews since they contain also the original results. This is relevant
especially for the papers \cite{3}, \cite{8} and \cite{13}. For example, the first
construction of the general non oscillatory solution with power asymptotic
near singularity for the case of perfect liquid with stiff matter equation of
state and investigation of the general oscillatory regime in presence of the
SU(2) Yang-Mills field can be found only in \cite{3}. The Iwasawa decomposition for
the metric tensor was first introduced namely in \cite{8}. Many original details of
the dynamical system approach to cosmological evolution can be found in \cite{13}.
The papers \cite{13} and \cite{14} we mentioned for completeness in order to remind to a
reader on the existence of an alternative approach to the analysis of the
character of cosmological singularity based on the representation of the
Einstein equations in the form of dynamical system and the serch for the
description of its attractor in vicinity to the singularity. This is powerful
method which can be considered as dual to the "cosmological billird approch"
to which the present talk is mainly dedicated. The restricted time for the
talk prevented to include a discussion also on this important aspect of the
theory.} covering also the contemporary development are \cite{1}-\cite{14}. The BKL
conjecture has its foundation in a collection of results and, first of all, it
asserts that the general solution containing the cosmological singularity
exists. This fundamental question of existence of such solution was the
principal goal of our work, however, we succeeded also in describing the
analytical structure of gravitational and matter fields in asymptotic vicinity
to the singularity and we showed that in most general physical settings such
solution has complicated oscillatory behaviour of chaotic character.

In order to avoid misunderstandings let's stress that under cosmological
singularity we mean the singularity in time, when singular manifold is
space-like, and when the curvature invariants together with invariant
characteristics of matter fields (like energy density) diverge on this
manifold. An intuitive feeling that there are no reasons to doubt in existence
of the general solution with cosmological singularity we have already in 1964
but another five years passed before the concrete structure have been
discovered. In 1965 appeared the important theorem of Roger Penrose \cite{15},
saying that under some conditions the appearance of incomplete geodesics in
space-time is unavoidable. This is also singularity but of different type
since, in general, incompleteness does not means that invariants diverge. In
addition the theorem can say nothing about the analytical structure of the
fields near the points where geodesics terminate. Then Penrose's result was
not of a direct help for us, nevertheless it stimulated our search. Today it
is reasonable to consider that the BKL conjecture and Penrose theorem
represent two sides of the phenomenon but the links are still far to be
understandable. This is because BKL approach deal with asymptotic in the
vicinity to the singularity and Penrose theorem has to do with global space-time.

It is worth to stress that some misleading statements are to be found in the
literature in relation to the aforementioned results. First of all, from the
BKL theory as well as from Penrose theorem not yet follows that cosmological
singularity is inevitable in General Relativity. BKL showed that the general
solution containing such singularity exists but general in the sense that
initial data under which the cosmological singularity is bound to appear
represent a set of nonzero measure in the space of all possible data. However,
we don't know \textquotedblleft how big\textquotedblright\ this measure is and
we have no proof that this set can cover the totality of initial data. In the
non-linear system can be many general solutions (that is, each containing
maximal number of arbitrary functional parameters) of different types
including also a general solution without singularity. Moreover there is the
proof \cite{16} of the global stability of Minkowski spacetime which means that at
least in some small (but finite) neighbourhood around it exists a general
solution without any singularity at any time. The same is true in relation to
the all versions of Penrose theorem: for these theorems to be applicable the
nontrivial initial/boundary conditions are strictly essential to be satisfied,
but an infinity of solutions can exists which do not meet such
conditions. The thorough investigation of applicability of Penrose theorem as
well as all its subsequent variations the reader can find in \cite{17,18}.

The second delusion is that the general solution with singularity can be
equally applied both to the singularity in future (Big Crunch) and to the
singularity in past (Big Bang) ignoring the fact that these two situations are
quite different physically. To describe what are going near cosmological Big
Crunch (as well as near the final stage of gravitational collapse of an
isolated object in its co-moving system) one really need the general solution
since in the course of evolution inescapably will arise the arbitrary
perturbations and these will reorganize any regime into the general one. The
Big Bang is not the same phenomenon. We don't know initial conditions at the
instant of singularity in principle and there are no reasons to expect that
they should be taken to be arbitrary. For example, we can not ruled out the
possibility that the Universe started exactly with the aid of the Friedmann
solution and it may be true that this does not means any fine tuning from the
point of view of the still unknown physics near such exotic state. Of course,
the arbitrary perturbations familiar from the present day physics will appear
after Big Bang but this is another story. The conclusion is that if somebody
found the general cosmological solution this not yet means that he knows how
Universe really started, however he has grounds to think that he knows at
least something about its end.

Sometimes one can find in literature the statement that in the BKL approach
only the time derivatives are important near singularity and because of this
the asymptotic form of Einstein equations became the ordinary differential
equations with respect to time. Such statement is a little bit misleading
since space-like gradients play the crucial role in appearing the oscillatory
regime. One of the main technical advantage of the BKL approach consists in
identification among the huge number of the space gradients those terms which
are of the same importance as time derivatives. In the vicinity to the
singularity these terms in no way are negligible, they act during the whole
course of evolution (although from time to time and during comparatively short
periods) and namely due to them oscillations arise. The subtle point here is
that asymptotically these terms can be represented as products of effective
scale coefficients, governing the time evolution of the metric, and some
factors containing space-like derivatives. This nontrivial separation
springing up in the vicinity to the singular point produce gravitational
equation of motion which effectively are the ordinary differential equations
in time because all factors containing space-like derivatives enter these
equations solely as external parameters, though dynamically influent
parameters. Owing to these ordinary differential equations the asymptotic
evolution can be described as motion of a particle in some external potential.
The aforementioned dominating space gradients create the reflecting potential
walls responsible for the oscillatory regime. For the case of homogeneous
cosmological model of the Bianchi IX type such potential have been described
by Misner \cite{19}. The literal assertion that \textquotedblleft only the time
derivatives are important near singularity\textquotedblright\ is correct just
for those cases when the general solution is of non oscillatory character and
has simple power asymptotic near singularity as, for instance, for the cases
of perfect liquid with stiff-matter equation of state \cite{3,20,21}, pure gravity
in space-time of dimension more than ten \cite{22}, or some other classes of
"subcritical" field models \cite{23}.

\section{Basic structure of cosmological singularity}

The character of the general cosmological solution in the vicinity to the
singularity can most conveniently be described in the synchronous reference
system, where the interval is of the form%

\begin{equation}
-ds^{2}=-dt^{2}+g_{\alpha\beta}dx^{\alpha}dx^{\beta}\label{1}%
\end{equation}

We use a system of units where the Einstein gravitational constant and the
velocity of light are equal to unity. The Greek indices refer to
three-dimensional space and assume the values 1,2,3. Latin indices $i,k$ will
refer to four-dimensional space-time and will take the values 0,1,2,3. The
coordinates are designated as $(x^{0},x^{1},x^{2},x^{3})=(t,x,y,z)$.

The Einstein equations in this reference system take the form%

\begin{equation}
R_{0}^{0}=\frac{1}{2}\dot{\kappa}+\frac{1}{4}\kappa_{\beta}^{\alpha}%
\kappa_{\alpha}^{\beta}=T\text{ }_{0}^{0}-\frac{1}{2}T\text{ },\label{2}%
\end{equation}

\begin{equation}
R_{\alpha}^{0}=\frac{1}{2}(\kappa_{,\alpha}-\kappa_{\alpha;\beta}^{\beta
})=T\text{ }_{\alpha}^{0}\text{ },\label{3}%
\end{equation}

\begin{equation}
R_{\alpha}^{\beta}=\frac{1}{2\sqrt{g}}(\sqrt{g}\kappa_{\alpha}^{\beta}\dot
{)}+P_{\alpha}^{\beta}=T\text{ }_{\alpha}^{\beta}-\frac{1}{2}\delta_{\alpha
}^{\beta}T\text{ },\label{4}%
\end{equation}
where the dot signifies differentiation with respect to time $t$ and%

\begin{equation}
\kappa_{\alpha\beta}=\dot{g}_{\alpha\beta}\text{ },\text{ \ \ \ \ }g=\det
g_{\alpha\beta}\text{ }.\label{5}%
\end{equation}
The tensorial operations on the Greek indices, as well as covariant
differentiation in this system are performed with respect to the
three-dimensional metric $g_{\alpha\beta}$. The quantity $\kappa$\ is a
three-dimensional contraction:%

\begin{equation}
\kappa=\kappa_{\alpha}^{\alpha}=(\ln g\dot{)}\text{ }.\label{6}%
\end{equation}
$P_{\alpha}^{\beta}$ is a three-dimensional Ricci tensor, expressed in terms
of $g_{\alpha\beta}$ in the same way as $R_{i}^{k}$ is expressed in terms of
$g_{ik}$. The quantities $T$ $_{0}^{0},T$ $_{\alpha}^{0}$ and $T$ $_{\alpha
}^{\beta}$ are components of the energy-momentum tensor $T$ $_{i}^{k}$
four-dimensional contraction of which is designated by $T$
.\ \ \ \ \ \ \ \ \ \ \
\begin{equation}
T\text{ =}T_{k}^{k}=T\text{ }_{0}^{0}+T\text{ }_{\alpha}^{\alpha}\text{
}.\label{7}%
\end{equation}
\qquad

It turn out that the general cosmological solution of Eqs. (2)-(4) in the
asymptotic vicinity of a singularity with respect to time is of an oscillatory
nature and may be described by an infinite alternation of the so-called Kasner
epochs. The notions of a Kasner epoch and of the succession of two of these
epochs are the key elements in the dynamics of the oscillatory regime. It is
most convenient to study their properties in the example of empty space, when
$T$ $_{i}^{k}=0$ and then take into account all the changes that may be
observed in the presence of matter. This procedure is reasonable since, in
general, the influence of matter upon the solution in the vicinity of the
singularity appears to be either negligible or can be put under the control.

So let's assume \ that the tensor $T$ $_{i}^{k}$ in Eqs.(2)-(4) equals zero. A
Kasner epoch is a time interval during which in Eq. (4) the three-dimensional
Ricci tensor \ $P_{\alpha}^{\beta}$\ may be neglected in comparison with the
terms involving time differentiation. Then from (2) and (4) we obtain the
following equations in this approximation:%

\begin{equation}
(\sqrt{g}\kappa_{\alpha}^{\beta}\dot{)}=0\text{ },\text{ \ \ \ \ \ \ }%
\dot{\kappa}+\frac{1}{2}\kappa_{\beta}^{\alpha}\kappa_{\alpha}^{\beta}=0\text{
}.\label{8}%
\end{equation}
Here and elsewhere we shall assert that the singularity corresponds to the
instant $t=0$ and we shall follow the evolution of the solution towards the
singularity, i.e., the variation of time as it decreases from certain values
$t>0$ down to $t=0$. Eq. (3) in the general case is of no interest for the
dynamics of the solution, since its role is reduced to the establishment of
certain additional relations on arbitrary three-dimensional functions
resulting from the integration of the Eqs.(2),(4) (that is of supplementary
conditions for initial data).

The general solution of Eqs. (8) may be written down in the form%

\begin{equation}
g_{\alpha\beta}=\eta_{AB}l_{\alpha}^{A}l_{\beta}^{B},\text{ \ \ \ }\eta
_{AB}=diag(t^{2p_{1}},t^{2p_{2}},t^{2p_{3}})\label{9}%
\end{equation}
where by the big Latin letters $A,B,C$ we designate the three-dimensional
frame indices (they take the values 1,2,3). The exponents $p_{1},p_{2},p_{3}$
and vectors $l_{\alpha}^{A}$ are arbitrary functions of the three-dimensional
coordinates $x^{\alpha}$. We call the directions along $l_{\alpha}^{A}$ as
Kasner axis, the triad $l_{\alpha}^{A}$ represents the common eigenvectors
both for metric $g_{\alpha\beta}$ and second form $\kappa_{\alpha\beta}$. The
exponents $p_{1},p_{2},p_{3}$ satisfy two relations:%

\begin{equation}
p_{1}+p_{2}+p_{3}=1,\text{ \ \ }p_{1}^{2}+p_{2}^{2}+p_{3}^{2}=1.\label{10}%
\end{equation}

It ensues from these relations that one of the exponents $p_{A}$ is always
negative while the two others are positive. Consequently, the space expands in
one direction and contracts in two others. Then the value of any
three-dimensional volume element decreases since, according to (9)-(10), the
determinant of $g_{\alpha\beta}$ decreases proportionally to $t^{2}$.

Of course, the solution (9)-(10) sooner or later will cease to be valid
because the\textit{\ }three-dimensional Ricci tensor\textit{\ \ }$P_{\alpha
}^{\beta}$ \ contain some terms which are growing with decreasing of time
faster than the terms with time derivatives and our assumption that
$P_{\alpha}^{\beta}$ can be neglected will become wrong. It is possible to
identify these "dangerous" terms in $P_{\alpha}^{\beta}$ and include them into
the new first approximation to the Einstein equations, instead of (8). The
remarkable fact is that the asymptotic solution of this new approximate system
can be described in full details and this description is valid and stable up
to the singularity. The result is that the evolution to the singularity can be
represented by a never-ending sequence of Kasner epochs and the singularity
$t=0$ is the point of its condensation. The durations of epochs tend to zero
and transitions between them are very short comparatively to its durations.
The determinant of the metric tensor $g_{\alpha\beta}$ tends to zero. On each
Kasner epoch the solution take the form (9)-(10) but each time with new
functional parameters $\acute{p}_{A}$ and $\acute{l}_{\alpha}^{A}$. On each
epoch the exponents $\acute{p}_{A}$ satisfy the same relations (10), that is
the space expands in one direction and contracts in two others, however, from
epoch to epoch these directions are different, i.e. on each new epoch the
Kasner axis rotate relatively to their arrangement at the preceding one.

The effect of rotation of Kasner axis make its use inconvenient for an
analytical description of the asymptotic oscillatory regime because this
rotation never stops. However, it turn out \ that another axis exist (they are
not eigenvectors for the second form $\kappa_{\alpha\beta}$) , rotation of
which are coming to stop in the limit $t\rightarrow0$ and projection of the
metric tensor into such "asymptotically frozen" (terminology of the authors of
Ref. \cite{8}) triad still is a diagonal matrix. The components of this matrix have
no limit since their behaviour again can be described by the never-ending
oscillations of a particle against some potential walls. This is an efficient
way to reduce the description of asymptotic evolution of six components of the
metric tensor to the three oscillating degrees of freedom. For the homogeneous
model of Bianchi type IX this approach was developed in \cite{24,25} where the
three-dimensional interval has been represented in the form $\ g_{\alpha\beta
}^{{}}dx^{\alpha}dx^{\beta}=(\tilde{R}\Gamma R)_{AB}(l_{\alpha}^{A}dx^{\alpha
})(l_{\beta}^{B}dx^{\beta})$ with the standard Bianchi IX differential forms
$l_{\alpha}^{A}dx^{\alpha}$ (where $l_{\alpha}^{A}$ depends only on
$x^{\alpha}$ in that special way that $\partial_{\nu}l_{\mu}^{C}-\partial
_{\mu}l_{\nu}^{C}=C_{AB}^{C}l_{\mu}^{A}l_{\nu}^{B}$ with only non-vanishing
structural constants $C_{23}^{1}=C_{31}^{2}=C_{12}^{3}=1$). The diagonal
matrix $\Gamma$ and three-dimensional \textit{orthogonal }matrix $R$ depend
only on time (tilde means transposition). Remarkably, the gravitational
equations for this model shows that near singularity all three Euler angles of
matrix $R$ \ tends to some arbitrary limiting constants and three components
of $\Gamma$ oscillate between the walls of potential of some special structure.

We never tried to generalize this approach (namely with orthogonal matrix $R$)
to the inhomogeneous models but the recent development of the theory showed
that even in most general inhomogeneous cases (including multidimensional
spacetime filled by different kind of matter) there is analogous
representation of the metric tensor leading to the same asymptotic freezing
phenomenon of "non-diagonal" degrees of freedom and reducing the full dynamics
to the few "diagonal" oscillating scale factors. This is so-called Iwasawa
decomposition first used in \cite{8} and thoroughly investigated in \cite{26}. The
difference is that in general inhomogeneous case instead of orthogonal matrix
$R$ it is more convenient to use an upper triangular matrix $N$ \ (with
components $N_{\alpha}^{A}$ where upper index $A$ numerates the rows and lower
index $\alpha$ corresponds to columns)%

\begin{equation}
N=\left(
\begin{array}
[c]{ccc}%
1 & n_{1} & n_{2}\\
0 & 1 & n_{3}\\
0 & 0 & 1
\end{array}
\right)  \text{\ \ }\label{11}%
\end{equation}
and to write three-dimensional interval in the form $g_{\alpha\beta}^{{}%
}dx^{\alpha}dx^{\beta}=(\tilde{N}\Gamma N)_{\alpha\beta}dx^{\alpha}dx^{\beta
})$. The diagonal matrix $\Gamma$ as well as matrix $N$ are functions of all
four coordinates but near the singularity matrix $N$ \ tends to some
time-independent limit and components of $\Gamma$ oscillate between the walls
of some potential. This asymptotic oscillatory regime has the well defined
Lagrangian. If one writes matrix $\Gamma$ as $\Gamma=diag(e^{-2\beta_{{}}^{1}%
},e^{-2\beta_{{}}^{2}},e^{-2\beta_{{}}^{3}})$ then the asymptotic equations of
motion for the scale coefficients $\beta_{\text{ }}^{A}$ became the ordinary
differential equations in time (separately for each point $x^{\alpha\text{ }}%
$of three-dimensional space) which follow from the Lagrangian:
\begin{equation}%
\begin{array}
[c]{ll}%
L=G_{AB}\frac{d\beta^{A}}{d\tau}\frac{d\beta^{B}}{d\tau}-V(\beta^{A}), & \\
V(\beta^{A})=C_{1}e^{-4\beta_{{}}^{1}}+\ C_{2}e^{-2(\beta_{{}}^{2}-\beta
^{1})^{{}}}+C_{3}e^{-2(\beta_{{}}^{3}-\beta^{2})}\ .\label{12} &
\end{array}
\end{equation}
Here we use the new time variable $\tau$ instead of original synchronous time
$t$. In asymptotic vicinity to the singularity the link is $dt=-\sqrt{\det
g_{\alpha\beta}}d\tau$ where differentials should be understood only with
respect to time, considering the coordinates $x^{\alpha}$ in $\det
g_{\alpha\beta}$ formally as fixed quantities. Since $\det g_{\alpha\beta}$
tends to zero approximately like $t^{2}$ it follows that singular limit
$t\rightarrow0$ corresponds to $\tau\rightarrow\infty.$ The metric $G_{AB} $
of three-dimensional space of scale coefficients $\beta_{\text{ }}^{A}$ are
defined by the relation $G_{AB}d\beta_{\text{ }}^{A}d\beta_{\text{ }}^{B}%
=\sum(d\beta_{\text{ }}^{A})^{2}-(\sum d\beta_{\text{ }}^{A})^{2}.$ This is
flat Lorenzian metric with signature (-,+,+) which can be seen from
transformation $\beta^{1}=\acute{\beta}^{1}+\acute{\beta}^{2}+\acute{\beta
}^{3},$ $\beta^{2}=\acute{\beta}^{1}-\acute{\beta}^{2}+\acute{\beta}^{3},$
$\beta^{3}=-\acute{\beta}^{3}$ \ after which one get $G_{AB}d\beta_{\text{ }%
}^{A}d\beta_{\text{ }}^{B}=-2(d\acute{\beta}^{1})^{2}+2(d\acute{\beta}%
^{2})^{2}+2(d\acute{\beta}^{3})^{2}.$ All coefficients $C_{A}(x^{\alpha})$ are
time-independent and positive; with respect to the dynamics they play a role
of external fixed parameters. Apart from the three differential equations of
second order for $\beta_{\text{ }}^{A}$ which follow from the Lagrangian (12)
there is well known additional constraint%
\begin{equation}
G_{AB}\frac{d\beta^{A}}{d\tau}\frac{d\beta^{B}}{d\tau}+V(\beta^{A}%
)=0,\label{13}%
\end{equation}
which represents the $(_{0}^{0})$ component of the Einstein equations. In
particular case of homogeneous model of Bianchi type IX equations (12) and
(13) gives exactly the same system which was described in \cite{24,25}, in spite of
the fact that in these last papers the asymptotical freezing of "non-diagonal"
metric components has been obtained using an orthogonal matrix $R$ instead of
Iwasawa's one $N.$ Analysis of the eqs. (12)-(13) shows that in the limit
$\tau\rightarrow\infty$ the exponents\ $\beta^{A}(\tau)$ are positive and all
tend to\ infinity in such a way that the differences $\beta^{2}-\beta^{1}$ and
$\beta^{3}-\beta^{2}$ also are positive and tend to infinity, that is each
term in the potential $V(\beta^{A})$ tends to zero. Then from (13) follows
that each trajectory $\beta^{A}(\tau)$ becomes "time-like" with respect \ to
the metric $G_{AB}$, i.e. near singularity we have $G_{AB}\frac{d\beta^{A}%
}{d\tau}\frac{d\beta^{B}}{d\tau}<0$, though this is so only in the extreme
vicinity to the potential walls $\beta^{1}=0,$ $\beta^{2}-\beta^{1}=0$ and
$\beta^{3}-\beta^{2}=0. $ Between the walls where $\beta^{1}>0,$ $\beta
^{2}-\beta^{1}>0,$ $\beta^{3}-\beta^{2}>0$ the potential is exponentially
small and trajectories become "light-like", i.e. $G_{AB}\frac{d\beta^{A}%
}{d\tau}\frac{d\beta^{B}}{d\tau}=0.$ These periods of "light-like" motion
between the walls corresponds exactly to the Kasner epochs (9)-(10) \ (with an
appropriate identification of Kasner axis during each period). It is easy to
see that the walls itself are "time-like" what means that collisions of a
"particle moving in a light-like directions" against the walls are inescapable
and interminable.

One of the crucial points discovered in \cite{8} is that in the limit
$\tau\rightarrow\infty$ the walls become infinitely sharp and of infinite
height which simplify further the asymptotic picture and make transparent the
reasons of chaoticity of such oscillatory dynamics. Because $G_{AB}%
\beta_{\text{ }}^{A}\beta_{\text{ }}^{B}=-2(\beta^{1}\beta^{2}+\beta^{1}%
\beta^{3}+\beta^{2}\beta^{3})$ and near singularity all $\beta_{\text{ }}^{A}$
are positive we have $G_{AB}\beta_{\text{ }}^{A}\beta_{\text{ }}^{B}<0$. Then
by the transformation $\beta_{\text{ }}^{A}=\rho\gamma^{A}$ one can introduce
instead of $\beta_{\text{ }}^{A}$ the "radial" coordinate $\rho>0 $
($\rho\rightarrow\infty$ when $\tau\rightarrow\infty$) and "angular"
coordinates $\gamma^{A}$ subjected to the restriction $G_{AB}\gamma_{\text{ }%
}^{A}\gamma_{\text{ }}^{B}=-1.$ The last condition pick out $\ $in $\gamma
$-space the two-dimensional Lobachevsky surface of constant negative curvature
and each trajectory $\beta^{A}(\tau)$ has the radially projected trace on this
surface. The free Kasner flights\ in three-dimensional $\beta$-space between
the walls are projected into geodesics of this two-dimensional surface. The
walls are projected into three curves forming a triangle on the Lobachevsky
surface and reflections against these curves are geometric (specular). If we
introduce the new evolution parameter $T $ by the relation $d\tau=\rho^{2}dT$
then the new Lagrangian (with respect to the ``time'' $T$) will be:
\begin{equation}%
\begin{array}
[c]{ll}%
L_{T}=-\left(  \frac{d\ln\rho}{dT}\right)  ^{2}+G_{AB}\text{ }\frac
{d\gamma^{A}}{dT}\frac{d\gamma^{B}}{dT}-\rho^{2}V(\rho\gamma^{A}), & \\
G_{AB}\gamma_{\text{ }}^{A}\gamma_{\text{ }}^{B}=-1.\text{\ \ }\label{14} &
\end{array}
\end{equation}

In the limit $\rho=\infty$ the new potential $\rho^{2}V(\rho\gamma^{A})$ is
exactly zero in the region between the walls where $\gamma^{1}>0,$ $\gamma
^{2}-\gamma^{1}>0,$ $\gamma^{3}-\gamma^{2}>0$ \ and becomes infinitely large
at the points $\gamma^{1}=0,$ $\gamma^{2}-\gamma^{1}=0,$ $\gamma^{3}%
-\gamma^{2}=0$ \ where the walls are located and behind them where quantities
$\gamma^{1},$ $\gamma^{2}-\gamma^{1},\gamma^{3}-\gamma^{2}$ are negative. This
means that near singularity potential $\rho^{2}V$ \ depends only on $\gamma
$-variables and $\rho$ can be considered as cyclic degree of freedom. In this
way the asymptotic oscillatory regime can be viewed as the eternal motion of a
particle inside a triangular bounded by the three stationary walls of infinite
height in two-dimensional space of constant negative curvature. The important
fact is that the area occupied by this triangle is finite. It is well known
(see references in \cite{8}, section 5.2.2) that the geodesic motion under the
conditions described is chaotic.

It is worth to mention that in case of homogeneous Bianchi IX model the fact
that its dynamics is equivalent to a billiard on the Lobachevsky plane was
established in \cite{27}.

The numerical calculations confirming the admissibility of the BKL conjecture
can be found in \cite{6} and \cite{28,29}.

\section{The influence of matter}

In papers \cite{3,20,30} we studied the problem of the influence of various kinds
of matter upon the behaviour of the general solution of the gravitational
equations in the neighbourhood of a singular point. It is clear that,
depending on the form of the energy-momentum tensor, we may meet three
different possibilities: (i) the oscillatory regime remains as it is in
vacuum, i.e. the influence of matter may be ignored in the first
approximation; (ii) the presence of matter makes the existence of Kasner
epochs near a singular point impossible; (iii) Kasner epochs exist as before,
but matter strongly affects the process of their formation and alternation.
Actually, all these possibilities may be realized.

There is a case in which the oscillatory regime observed as a singular point
is approached remains the same, in the first approximation, as in vacuum. This
case is realized in a space filled with a perfect liquid with the equation of
state $p=k\varepsilon$ for $0<k<1$. No additional reflecting walls arise from
the energy-momentum tensor in this case.

If $k=1$ we have the "stiff matter" equation of state $p=\varepsilon$. This is
the second of the above-mentioned possibilities when neither Kasner epoch nor
oscillatory regime can exist in the vicinity of a singular point. This case
has been investigated in \cite{3,20} where it has been shown that the influence of
the "stiff matter" (equivalent to the massless scalar field) results in the
violation of the Kasner relations (10) for the asymptotic exponents. Instead
we have%
\begin{equation}
p_{1}+p_{2}+p_{3}=1,\text{ \ \ \ \ }p_{1}^{2}+p_{2}^{2}+p_{3}^{2}%
=1-p_{\varphi}^{2}\label{15}%
\end{equation}
where $p_{\varphi}^{2}$ is an arbitrary three-dimensional function (with the
restriction $p_{\varphi}^{2}<1$) to which the energy density $\varepsilon$ of
the matter is proportional (in that particular case when the stiff-matter
source is realized as a massless scalar field $\varphi$ its asymptotic is
$\varphi=p_{\varphi}\ln t$ and this is the formal reason why we use the index
$\varphi$ for the additional exponent $p_{\varphi}$).

Thanks to (15), in contrast to the Kasner relations (10), it is possible for
all three exponents $p_{A}$ to be positive. In \cite{20} it has been shown that,
even if the contraction of space starts with the quasi-Kasner epoch (15)
during which one of the exponents $p_{A}$ is negative, the asymptotic
behaviour (9) with all positive exponents is inevitably established after a
finite number of oscillations and remains unchanged up to the singular point.
Thus, for the equation of state $p=\varepsilon$ the collapse in the general
solution is described by monotonic (but anisotropic) contraction of space
along all directions. The asymptotic of the general solution near cosmological
singularity for this case we constructed explicitly in \cite{3}, see also
\cite{20,21,23}. The disappearance of oscillations for the case of a massless
scalar field should be consider as an isolated phenomenon which is unstable
with respect to inclusion into the right hand side of the Einstein equations
another kind of fields. For instance, in the same paper \cite{20} we showed that if
to the scalar field we add a vector one then the endless oscillations reappear.

The cosmological evolution in the presence of an electromagnetic field may
serve as an example of the third possibility. In this case the oscillatory
regime in the presence of matter is, as usual, described by the alternation of
Kasner epochs, but in this process the energy-momentum tensor plays a role as
important as the three-dimensional curvature tensor. This problem has been
treated by us in \cite{30}, where it has been shown that in addition to the vacuum
reflecting walls also the new walls arise caused by the energy-momentum tensor
of the electromagnetic field. The electromagnetic type of alternation of
epochs, however, qualitatively takes place according to the same laws as in vacuum.

In paper \cite{3} we have also studied the problem of the influence of the
Yang-Mills fields on the character of the cosmological singularity. For
definiteness, we have restricted ourselves to fields corresponding to the
gauge group SU(2). The study was performed in the synchronous reference system
in the gauge when the time components of all three vector fields are equal to
zero. It was shown that, in the neighbourhood of a cosmological singularity,
the behaviour of the Yang-Mills fields is largely similar to the behaviour of
the electromagnetic field: as before, there appears an oscillatory regime
described by the alternation of Kasner epochs, which is caused either by the
three-dimensional curvature or by the energy-momentum tensor. If, in the
process of alternation of epochs, the energy-momentum tensor of the gauge
fields is dominating, the qualitative behaviour of the solution during the
epochs and in the transition region between them is like the behavior in the
case of free Yang-Mills fields (with the Abelian group). This does not mean
that non-linear terms of the interaction may be neglected completely, but the
latter introduce only minor, unimportant quantitative changes into the picture
we would observe in the case of non-interacting fields. The reason for this
lies in the absence of time derivatives of the gauge field strengths in those
terms of the equations of motion which describe the interaction.

\section{Multidimensional spacetime and supergravity}

The story resembling the aforementioned effect of dissappearance (for scalar
field) and reconstruction (after adding a vector field) of oscillations
occurred later in more general and quite different circumstances. In 1985
appeared very interesting and unexpected result \cite{22} that oscillatory regime
near cosmological singularity in multidimensional spacetime (for pure gravity)
holds for spacetime dimension $D$ up to $D=10$ but for dimension $D\geq11$ the
asymptotic of the general solution follow the smooth multidimensional Kasner
power law. Up to now we have no idea why this separating border coincides with
dimension so significant for superstring theories, most likely it is just an
accident. However, the important point is that if we will add to the vacuum
multidimensional gravity the fields of $p$-forms the presence of which is
dictated by the low energy limit of superstring models, the oscillatory regime
will reappear. This fact was established in \cite{31,32} and subsequently has been
developed by T.Damour, M.Henneaux, H.Nicolai, B.Julia and their collaborates
into the new interesting and promising branch of superstring theories. In
articles \cite{31,32} it was demonstrated that bosonic sectors of supergravities
emerging in the low energy limit from all types of superstring models
have\ oscillatory cosmological singularity of the BKL character. Let consider
the action of the following general form:%
\begin{equation}%
\begin{array}
[c]{ll}%
\displaystyle S=\int d^{D}x\sqrt{g}\text{ }\biggl[R-\partial^{i}%
\varphi\partial_{i}\varphi- & \\
\displaystyle-\frac{1}{2}\sum_{p}\frac{1}{(p+1)!}e^{\lambda_{p}\varphi
}F_{i_{1}...i_{p+1}}^{(p+1)}F^{(p+1)i_{1}...i_{p+1}}\biggr]\label{16} &
\end{array}
\end{equation}
where $F_{{}}^{(p+1)}$ designates the the field strengths generated by the $p
$-forms $A_{p}$, i.e. $F_{i_{1}...i_{p+1}}^{(p+1)}=antisym(\partial_{i_{1}%
}A_{i_{2}...i_{p+1}})$. The real parameters $\lambda_{p}$ are coupling
constants corresponding to the interaction between the dilaton and $p$-forms.
The tensorial operations in (16) are carring out with respect to $D
$-dimensional metric $g_{ik}$ and $g=\left\vert \det g_{ik}\right\vert $. Now
the small Latin indices refer to $D$-dimensional space-time and Greek indices
(as well as big Latin frame indices $A,B$ and $C$) correspond to
$d$-dimensional space where $d=D-1.$ Also in this theory the Kasner-like
epochs exist which are of the form:
\begin{equation}%
\begin{array}
[c]{ll}%
g_{ik}dx^{i}dx^{k}=-dt^{2}+\eta_{AB}(t,x^{\alpha})l_{\mu}^{A}(x^{\alpha
})l_{\nu}^{B}(x^{\alpha})dx^{\mu}dx^{\nu}, & \\
\eta_{AB}=diag[t^{2p_{1}(x^{\alpha})},t^{2p_{2}(x^{\alpha})},...,t^{2p_{d}%
(x^{\alpha})}],\text{ \ \ }\label{17} &
\end{array}
\end{equation}%
\begin{equation}
\varphi=p_{\varphi}(x^{\alpha})\ln t+\varphi_{0}(x^{\alpha}).\label{18}%
\end{equation}
However, in the presence of the dilaton the exponents $p_{A\text{ }}$ instead
of the Kasner law satisfy the relations analogous to (15):
\begin{equation}
\sum_{A=1}^{d}p_{A}=1,\text{ \ \ }\sum_{A=1}^{d}p_{A}^{2}{}^{{}}=1-p_{\varphi
}^{2}\label{19}%
\end{equation}

The approximate solution (17)-(19) follows from the $D$-dimensional Einstein
equations by neglecting the energy-momentum tensor of $p$-forms,
$d$-dimensional curvature tensor $P_{\alpha}^{\beta}$ and spatial derivatives
of $\varphi$. Now one has to do the work analogous to that one for
4-dimensional gravity: it is necessary to identify in all neglected parts of
the equations those "dangerous" terms which will destroy the solution
(17)-(19) in the limit $t\rightarrow0$. Then one should construct the new
first approximation to the equations taking into account also these
``dangerous'' terms and try to find asymptotic solution for this new system.
This is the same method which have been used in case of the 4-dimensional
gravity with electromagnetic field and it works well also here. Using the
Iwasawa decomposition for $d$-dimensional frame metric \ $\eta_{AB}=(\tilde
{N}\Gamma N)_{AB}$ where $\Gamma=diag(e^{-2\beta_{{}}^{1}},e^{-2\beta_{{}}%
^{2}},...,e^{-2\beta_{{}}^{d}})$ it can be shown \cite{8} that near singularity
again the phenomenon of freezing of "non-diagonal" degrees of freedom of the
metric tensor arise and the foregoing new approximate system reduces to the
ordinary differential equations (for each spatial point) for the variables
$\beta^{1},...,\beta^{d}$ and $\beta^{d+1}$ where $\beta^{d+1}=-\varphi.$ It
is convenient to use the $d+1$-dimensional flat superspace with coordinates
$\beta^{1},...,\beta^{d+1}$ and correspondingly new indices $\bar{A},\bar{B}$
running over values $1,...,d+1. $ The metric $G_{\bar{A}\bar{B}}$ in this
superspace is
\begin{equation}
G_{\bar{A}\bar{B}}d\beta_{\text{ }}^{\bar{A}}d\beta_{\text{ }}^{\bar{B}}%
=\sum_{A=1}^{d}(d\beta_{\text{ }}^{A})^{2}-(\sum_{A=1}^{d}d\beta_{\text{ }%
}^{A})^{2}+\left(  d\beta^{d+1}\right)  ^{2}.\label{20}%
\end{equation}
The asymptotic dynamics for $\beta$-variables follows from the Lagrangian of
the form similar to (14):%

\begin{equation}%
\begin{array}
[c]{ll}%
L_{T}=-\left(  \frac{d\ln\rho}{dT}\right)  ^{2}+G_{\bar{A}\bar{B}}\text{
}\frac{d\gamma^{\bar{A}}}{dT}\frac{d\gamma^{\bar{B}}}{dT}-\rho^{2}\sum
_{b}C_{b}e^{-2\rho w_{b}(\gamma)}, & \\
G_{\bar{A}\bar{B}}\gamma_{\text{ }}^{\bar{A}}\gamma_{\text{ }}^{\bar{B}%
}=-1.\label{21} &
\end{array}
\end{equation}
Again $(_{0}^{0})$ component of the Einstein equations gives additional
condition to the equations of motion following from this Lagrangian:%
\begin{equation}
-\left(  \frac{d\ln\rho}{dT}\right)  ^{2}+G_{\bar{A}\bar{B}}\text{ }%
\frac{d\gamma^{\bar{A}}}{dT}\frac{d\gamma^{\bar{B}}}{dT}+\rho^{2}\sum_{b}%
C_{b}e^{-2\rho w_{b}(\gamma)}=0.\label{22}%
\end{equation}
Here $\beta_{\text{ }}^{\bar{A}}=\rho\gamma^{\bar{A}}$ and time parameters $T
$ \ and $\tau$ are defined by the evident generalization to the
multidimensional spacetime of their definitions we used in case of
4-dimensional gravity: $dt=-\sqrt{\det g_{\alpha\beta}}d\tau,$ $d\tau=\rho
^{2}$$dT$ . All functional parameters $C_{b}(x^{\alpha})$ in general are
positive. The cosmological singularity corresponds to the the limit
$\rho\rightarrow\infty$ and in this limit potential term in Lagrangian can be
considered as $\rho$-independent, asymptotically it vanish in the region of
this space where $w_{b}(\gamma)>0$ and is infinite where $w_{b}(\gamma)<0.$The
sum in the potential means summation over all relevant (dominating)
impenetrable barriers located at hypersurfaces where $w_{b}(\gamma)=0$ in the
hyperbolic $d$-dimensional $\gamma$-space. All $w_{b}(\gamma)$ are linear
functions on $\gamma$ therefore $w_{b}(\gamma)=\rho^{-1}w_{b}(\beta).$ The
free motion of $\beta_{\text{ }}^{\bar{A}}(\tau)$ between the walls in the
original $d+1$-dimensional $\beta$-superspace is projected onto a geodesic
motion of $\gamma^{\bar{A}}(T)$ on hyperbolic $d$-dimensional $\gamma$-space,
i.e. to the motion between the corresponding projections of the original walls
onto $\gamma$-space. These geodesic motions from time to time are interrupted
by specular reflections against the infinitely sharp hyperplanes $w_{b}%
(\gamma)=0$. These hyperplanes bound a region in $\gamma$-space inside which a
symbolic particle oscillates and the volume of this region, in spite of its
non-compactness, is finite. The last property is of principle significance
since it leads to the chaotic character of the oscillatory regime.

Of course, one of the central point here is to find all the aforementioned
dominant walls and corresponding "wall forms" $w_{b}(\beta).$ This depends on
the spacetime dimension and menu of $p$-forms. In papers \cite{33,7,8} the
detailed description of all possibilities for the all types of supergravities
(i.e., eleven-dimensional supergravity and those following from the known five
types of the superstring models in ten-dimensional spacetime) can be found. It
was shown that in all cases there is only 10 relevant walls governing the
oscillatory dynamics. The large number of other walls need no consideration
because they are located behind these principal ten and have no influence on
the dynamics in the first approximation. The mentioned above region in
$\gamma$-space where a particle oscillate is called "billiard table" and
collection of its bounding walls forms the so-called Coxeter crystallographic
simplex, that is, in the cases under consideration, polyhedron with 10 faces
in 9-dimensional $\gamma$-space with all dihedral angles between the faces
equal to the numbers $\pi/n$ where $n$ belongs to some distinguished set of
natural numbers (or equal to infinity). This is very special geometric
construction which (when combined with the specular laws of reflections
against the faces) lead to the nontrivial huge symmetry hidden in the
asymptotic structure of spacetime near cosmological singularity which symmetry
coexists, nevertheless, with chaoticity.

\section{Damour-Hennaux-Nicolai hidden symmetry conjecture}

The mathematical description of the symmetry we are talking about can be
achieved in the following way. Consider the trajectories $\beta_{\text{ }%
}^{\bar{A}}(\tau)$ of a particle moving between the walls $w_{b}(\beta)=0$ in
the original 10-dimensional $\beta$-superspace with coordinates $\beta_{\text{
}}^{\bar{A}}$ and metric $G_{\bar{A}\bar{B}}$ (20). These trajectories are
null stright lines with respect to the Lorenzian metric $G_{\bar{A}\bar{B}}.$
Wall forms $w_{b}(\beta)$ are linear function on $\beta$, that is
$w_{b}=w_{b\bar{A}}\beta_{\text{ }}^{\bar{A}}$ where the set of constants
$w_{b\bar{A}}$ depends on the choice of a supergravity model and on the type
of the wall (index $b$) in the chosen model. \ We see that for each wall
$w_{b}=0$ the constants $w_{b\bar{A}}$\ represent components of the vector
orthogonal to to this wall. We can imagine all these vectors (for different
$b$) as arrows starting at the origin of the $\beta$-space. All these vectors
have fixed finite norm $G^{\bar{A}\bar{B}}w_{b\bar{A}}w_{b\bar{B}}$
($G^{\bar{A}\bar{B}}$ is inverse to \ $G_{\bar{A}\bar{B}}$) and one can
arrange the scalar products $(w_{a}\bullet w_{b})=G^{\bar{A}\bar{B}}%
w_{a\bar{A}}w_{b\bar{B}}$ for each supergravity model in the form of the
matrix:%
\begin{equation}
A_{ab}=2\frac{(w_{a}\bullet w_{b})}{(w_{a}\bullet w_{a})}\text{ \ \ (no
summation in }a\text{).}\label{23}%
\end{equation}
\ 

The crucial point is that, independently of a supergravity model, $A_{ab}$ is
the Cartan matrix of indefinite type, i.e. with one negative principal value
\cite{8,33,34,35}. Any Cartan matrix can be associated with some Lie algebra and
particular matrix (23) corresponds to the so-called Lorenzian hyperbolic
Kac-Moody algebra of the rank 10. As was shown in \cite{31} the particle's velocity
$v^{\bar{A}}=d\beta^{\bar{A}}/d\tau$ after the reflection from the wall
$w_{a\bar{A}}\beta_{\text{ }}^{\bar{A}}=0$ changes according to the universal
(i.e. again independent of the model) law:
\begin{equation}%
\begin{array}
[c]{ll}%
(v^{\bar{A}})_{after}=(v^{\bar{A}})_{before}-2\frac{(v^{\bar{B}}%
)_{before}w_{a\bar{B}}}{(w_{a}\bullet w_{a})}w_{a}^{\bar{A}}, & \\
w_{a}^{\bar{A}}=G^{\bar{A}\bar{B}}w_{a\bar{B}},\text{\ \ \ (no summation in
}a\text{).}\label{24} &
\end{array}
\end{equation}
This transformation is nothing else but the already mentioned specular
reflection of a particle by the wall orthogonal to the vector $w_{a\bar{A}}.$
Now it is clear that one can formally identify the ten vectors $w_{a\bar{A}}$
with the simple roots of the root system of Kac-Moody algebra, the walls
$w_{a\bar{A}}\beta_{\text{ }}^{\bar{A}}=0$ with the Weyl hyperplanes
orthogonal to the simple roots, the reflections (24) with the elements of the
Weyl group of the root system and the region of $\beta$-superspace bounded by
the walls (where a particle oscillates) with the fundamental Weyl chamber. For
the readers less familiar with all these notions of the theory of generalized
Lie algebras (especially in application to the question under consideration)
we can recommend the exhaustive review \cite{10} which is well written also from
pedagogical point of view.

The manifestation of Lie algebra means that the corresponding Lie symmetry
group must somehow be hidden in the system . The hidden symmetry conjecture
\cite{35,36,37} proposes that this symmetry might be inherent for the exact
superstring theories (assuming that they exist) and not only for their
classical low energy limits of their bosonic sectors in the vicinity to the
cosmological singularity. The limiting structure near singularity should be
considered just as an auxiliary instrument by means of which this symmetry is
coming to light. As of now we have no comprehension where and how exactly the
symmetry would act (could be as a continuous infinite dimensional symmetry
group of the exact Lagrangian permitting to transform the given solutions of
the equations of motion to the new solutions). If true the hidden symmetry
conjecture could create an impetus for the third revolution in the development
of the superstring theories.

\vspace{1cm}

{\bf Acknowledgments}

I would like to express the gratitude to the organizers of this conference
for the excellent arrangement and for the warm hospitality in Minsk. I am also
grateful to Thibault Damour for useful comments which helped me to improve
the present exposition of my talk.

\end{document}